\documentclass[12pts]{article}
\title{On the low temperature series expansion for the diagonal correlation functions in the two-dimensional Ising model}
\author{Ranjan Kumar Ghosh\thanks{\normalsize e-mail address:- rkg\_1978@yahoo.com}\\Bidhannagar College, EB-2 Salt Lake City,\\ Calcutta-\ 700 064, INDIA}
\date{May 2,2005}
\begin{document}
\Large
\maketitle
\begin{abstract}\normalsize
A previously tested differential equation method for generating low temperature series expansion for diagonal spin-spin correlation functions in the $d=2$ Ising model is extended to generate the non-universal terms for arbitrary separation of the spins.This extends the earlier calculations of these correlation functions.
\end{abstract}

The two-dimensional Ising model is one of the very few exactly soluble systems$^{[1]}$.Many thermodynamic functions for this system have been calculated(exactly or approximately) for this system in classic papers long time ago$^{[2,3]}$.Among the interesting objects of this model are the (static) spin-spin correlation functions\  $<\sigma_{00}\sigma_{mn}> \ ^{[4],[5],[6]}$.A few years ago, the author in a collaboration , used a differential equation approach to the diagonal correlation functions$\ S_n= <\sigma_{00}\sigma_{nn}> $ of this model obtaining the solutions as high and low temperature series expansions in suitable expansion variables$^{[7]}$. Although the high temperature expansions generated this way were very successful in getting the exact coefficients for arbitrary $n$,the low temperature expansion was not very successful as it was realised that these expansions have a universal part which comes from the series for the exact expression for the spontaneous magnetisation of the system,and the rest of the expansions have a non-universal character i.e. they depend on the particular value of $n$.In that paper we did not say anything about how to go about calculating these non-universal terms. Soon afterwards, the same authors calculated the exact expressions for these correlation functions for values of $n$ upto $n=6\ ^{[8]}$.  However these expressions become more and more complicated as the value of $\ n$ grows and because of this , these were given only for values of $\ n$ upto$\ n=5$ in the publication at that time.  It is clear that if we want to write down the expressions exactly for large values of $\ n$ the computation time and the labour involved starts becoming unmanageable.Hence it seems worthwhile to look for a method for calculating these non-universal terms in the low temperature series expansion. At the same time ,to the author's knowledge, there exists no systematic method for calculating the low temperature expansion for these correlation functions.  The purpose of this paper is to show how we can calculate these very terms by the same differential equation approach that we had presented earlier. For the exact coefficients for the high temperature series for these correlation functions the reader is referred to[7]. In this short communication we follow the conventions adopted in our earlier paper to avoid any confusion in the results.

We take the Hamiltonian for the model to be
 \begin{equation}H=-\sum_{i,j\in Z^2}  (J_1\sigma_{i,j}\sigma_{i+1,j}+J_2\sigma_{i,j}\sigma_{i,j+1})\end{equation}where$\ J_1$ and $\ J_2$ are the exchange constants of the model.  
Let$\beta=(k_BT)^{-1}$ and as usual define\begin{equation}\ k_>=sinh(2\beta J_1)sinh(2\beta J_2)\end{equation}\\applicable for\ $T>T_c$,and
\begin{equation}k_<=k_{>}^{-1}\end{equation}applicable for$\  T<T_c$,\  where \  $T_c$ is   the   critical   temperature \  of \  the \  model,\   which\   is \  given \  by\begin{equation}k_{>}(\beta_c)=k_{<}(\beta_c)=1\end{equation}

The analysis then proceeds as follows. The variable $t$ is defined as \\
\begin{equation}t=\left\{\begin{array}{c}
k_{>}^{-2} for\  T>T_{c}\\k_{<}^{-2} for\  T<T_{c}\end{array}\right.\end{equation}
If$\ S_{n,\mp}$ are defined to be$\ <\sigma_{00}\sigma_{nn}>$ for $T<T_{c}$ and $T>T_{c}$ respectively and we define two related functions by
\begin{equation}\sigma_{n,-}=t (t-1)\frac{d}{dt}lnS_{n,-}-\frac{1}{4},\end{equation}
\begin{equation}\sigma_{n,+}=t (t-1)\frac{d}{dt}lnS_{n,+}-\frac{1}{4}t.\end{equation}
Jimbo and Miwa showed that$\ \sigma_{n}=\sigma_{n,\pm}\left(t\right)$satisfy the ordinary differential equation$^{[9]}$
\begin{equation}\left[t(t-1)\sigma^{\prime\prime}_{n}\right]^2-n^2\left[(t-1)\sigma^{\prime}_{n}-\sigma_{n}\right]^2+4\sigma^{\prime}_{n}\left[(t-1)\sigma^{\prime}_{n}-\sigma_{n}-\frac{1}{4}\right](t\sigma^{\prime}_{n}-\sigma_{n})=0\end{equation}

It was shown in ref.[7] that if we consider the high temperature phase then we can generate the series expansion for the spin-spin correlation functions rather easily,as there are graphical methods available for these expansions which give us the first term in the series with a little bit of combinatorics and then many terms of the series can be generated recursively using the differential equation. The program was very successful when carried out this way. However when the same equation was applied to the low temperature phase of the theory for a series expansion,beginning with the first term as $1$(which is the appropriate term for zero temperature) it ran into a problem.The recursive procedure gave us an expansion
\begin{equation}S_{n,-}=\sum_{l=0}^{n} \left(-1\right)^l\left(\begin{array}{c}\frac{1}{4}\\l\end{array}\right)k^{2l}_{<}+O(k^{2l+2}_{<})\end{equation}
where $\left(\mu \atop\nu\right)$is the binomial coefficient  in the expansion of the square of spontaneous magnetisation, given by
\begin{equation}M^2=\lim_{m^2+n^2 \rightarrow \infty}<\sigma_{00}\sigma_{mn}>=(1-k^{2}_{<})^{\frac{1}{4}}\end{equation}

 and the terms of the order$\ k^{2l+2}_{<}$ in equation(9) depend on the specific value of $n$. Because of some reasons , the calculation of the $n$ dependent terms could not be carried out at that time. Although we think that they carry important information about the detailed nature of spin-spin correlations along the diagonal.

The reason for our failure at that time was the fact that once we take the series and substitute it into the equation the coefficients of the terms of the order$\ k^{(2l+2)}_{<}$ and$\ k^{(2l+4)}_{<}$ get coupled together in the same term along with the other coefficients.This makes it impossible to determine the first $n$ dependent term in the series without recourse to some other information.

Fortunately a closer examination after all these years has revealed that the first $n$ dependent term in the low temperature expansion of $S_{n,-}$is given by the expression
\begin{equation}C_{n,2n+2}=\left(-1\right)^{(n+1)}\left(\begin{array}{c}\frac{1}{4}\\n+1\end{array}\right)+\frac{1}{\left(2n+1\right)} \left(\frac{\left(2n+1\right)!!}{2^{(n+1)} (n+1)!}\right)^2\end{equation}Where $S_{n,-}$ is given by
\begin{equation}S_{n,-}\ =\ \sum_{m=0}^\infty\ C_{n,m}k_{<}^m\end{equation}

This relation
\footnote{\normalsize This relation seems to imply the equality\ $\left(\frac{d}{dk^2}\right)^{n+1}(S_{1,-}+S_{n,-})|_{k=0}=\left(\frac{d}{dk^2}\right)^{n+1} M^2|_{k=0}$.J.H.H. Perk has informed me that the proof of this equality is easy using the quadratic recurrence relations for the correlation functions given in the ref.[10].}
has been tested for $n\le 6$ and is conjectured to hold for arbitrary $n$.Armed with this information we go on to calculate the series for $S_{n,-}$ for values of $n$ upto $10$.It is seen that with this value of the coefficient, the series can be generated term by term, and this can not be done for the other possible values of the coefficient that we can try.However the most important criterion for choosing this function as the correct value of the above term is the fact that it gives the correct coefficients for all the subsequent terms of the series for known exact correlation functions. We carry out the expansion for the first thirty terms in series from $n=6$ to $n=10$. The expressions for these functions upto $n=5$ have been omitted because the exact expressions for these are rather simple in this case, and are already available in the literature.

\begin{equation}C_{n,m}\ =0,\  for\  m\ =\ odd\  no.\end{equation}

$n=6$\\
\begin{center}
$\begin{array}{lcl}C_{6,14}=\ -\frac{62755}{4194304}&\ & C_{6,16}=\ -\frac{5885649}{536870912}\\
\\
C_{6,18}=\ -\frac{18102051}{2147483648}&\ & C_{6,20}=\ -\frac{115143369}{17179869184}\\
\\
C_{6,22}=\ -\frac{1502308899}{274877906944}&\ & C_{6,24}=\ -\frac{39992200521}{8796093022208}\\
\\
C_{6,26}=\ -\frac{135253617001}{35184372088832}&\ & C_{6,28}=\ -\frac{927283251903}{281474976710656}\\
\\
C_{6,30}=\ -\frac{12860686137607}{4503599627370496}&\ &\ \end{array}$\\
\end{center}

$n=7$\\
\begin{center}
$\begin{array}{lcl}C_{7,16}=\ -\frac{13852377}{1073741824}& \ & C_{7,18}=\ -\frac{41988023}{4294967296}\\
\\
C_{7,20}=\ -\frac{529859407}{68719476736}& \ & C_{7,22}=\ -\frac{1719298113}{274877906944}\\
\\
C_{7,24}=\ -\frac{91202314827}{17592186044416}& \ & C_{7,26}=\ -\frac{307615754733}{70368744177664}\\
\\
C_{7,28}=\ -\frac{1052329791057}{281474976710656}& \ & C_{7,30}=\ -\frac{1821460877687}{562949953421312}\end{array}$\\
\end{center}

$n=8$\\
\begin{center}
$\begin{array}{lcl}C_{8,18}=\ -\frac{48531703}{4294967296}& \ & C_{8,20}=\ -\frac{302177557}{34359738368}\\
\\
C_{8,22}=\ -\frac{973272573}{137438953472}& \ & C_{8,24}=\ -\frac{25689376531}{4398046511104}\\
\\
C_{8,26}=\ -\frac{172699772293}{35184372088832}& \ & C_{8,28}=\ -\frac{1178557994739}{281474976710656}\\
\\
C_{8,30}=\ -\frac{4071839681233}{1125899906842624}& \ &\ \end{array}$\\
\end{center}

$n=9$\\
\begin{center}
$\begin{array}{lcl}C_{9,20}=\ -\frac{688829933}{68719476736}& \ & C_{9,22}=\ -\frac{2191660887}{274877906944}\\
\\
C_{9,24}=\ -\frac{114906443431}{17592186044416}& \ & C_{9,26}=\ -\frac{384494320401}{70368744177664}\\
\\
C_{9,28}=\ -\frac{1307717138979}{281474976710656}& \ & C_{9,30}=\ -\frac{281692820385}{70368744177664}\end{array}$\\
\end{center}

$n=10$\\
\begin{center}
$\begin{array}{lcl}C_{10,22}=\ -\frac{2470054587}{274877906944}& \ & C_{10,24}=\ -\frac{64020065109}{8796093022208}\\
\\
C_{10,26}=\ -\frac{212856341061}{35184372088832}& \ & C_{10,28}=\ -\frac{1441790227059}{281474976710656}\\
\\
C_{10,30}=\ -\frac{19816146315693}{4503599627370496}\end{array}$\\
\end{center}

We have stopped at the terms of the order $\ k_{<}^{30}$ in all these expansions. But if needed these series can be continued to higher powers of $k_{<}$.

Thus in this paper we have given a scheme of calculating the low temperature expansions of the diagonal correlation functions in the  {\em\ two-dimensional} Ising model and we hope that this method will be useful in getting information about similar quantities in other models for which one can write down differential equations.

I am thankful to J.H.H.Perk for his comments and suggestions on the first version of this note. I am also thankful to him for bringing to my notice several recent references including ref.[5],[6] and [10]. The ref.[5] is specially important, for the series for susceptibility has been calculated in there. The incompleteness of the list of references is, however, due to the author's unfamiliarity with the later works.

\end{document}